\documentclass{PoS}

\usepackage{graphicx}
\usepackage{amsmath,amssymb}
\newcommand{\ba}{\begin{eqnarray}}
\newcommand{\ea}{\end{eqnarray}}
\newcommand{\be} {\begin{equation}}
\newcommand{\ee} {\end{equation}}
\newcommand{\GeV}{\mbox{\rm GeV}}
\newcommand{\order}{O}

\title{A determination of the $B_s^0$ and $B_d^0$ mixing parameters 
in 2+1 lattice QCD}

\ShortTitle{A determination of the $B_s^0$ and $B_d^0$ mixing parameters 
in 2+1 lattice QCD}

\author{R. Todd Evans, \speaker{Elvira G\'amiz} and Aida X. El-Khadra\\
        Department of Physics, University of Illinois, Urbana, IL 61801, USA\\
        E-mail: \email{rtevans@uiuc.edu}, \email{megamiz@uiuc.edu}, 
        \email{axk@uiuc.edu} 
        }

\author{Massimo Di Pierro\\
        School of Computer Sci., Telecom. and Info. Systems, DePaul 
        University, Chicago, IL, USA\\
        \email{MDiPierro@cti.depaul.edu}
        }

\author{Fermilab Lattice and MILC Collaborations}

\abstract{   
We report on the advances in our unquenched calculation of the matrix
elements relevant for the analysis of $B^0-\bar B^0$ mixing using the Asqtad
(light quark) and Fermilab (heavy quark) actions. We
have calculated the hadronic parameters for the mass and width differences in
the neutral $B$ meson system. Preliminary results are presented for 
$f_{B_q}^2 B_q$ as well as for the ratio 
$\xi^2=f_{B_s}^2 B_{B_s}/f_{B_d}^2 B_{B_d}$.

}

\FullConference{The XXV International Symposium on Lattice Field Theory\\
		 July 30-4 August 2007\\
		 Regensburg, Germany}

\begin{document}

\section{Introduction}

\label{introduccion}

The very accurate experimental measurements of the mass differences  
between the heavy and light $B^0_s$ and $B^0_d$ mass eigenstates, 
$\Delta M_s$ \cite{Msdetermination} and $\Delta M_d$ \cite{Mddetermination}, 
that describe the $B^0_s-\bar B^0_s$ and $B^0_d-\bar B^0_d$ mixings 
respectively, make improving the theoretical study of these 
quantities crucial. 
In the standard model (SM), the mass difference is given by \cite{BJW90} 
\ba\label{SMMsd}
\Delta M_{s(d)}\vert_{theor.}=\frac{G_F^2M_W^2}{6\pi^2}
\vert V_{t s(d)}^*V_{tb}\vert^2
\eta_2^BS_0(x_t)M_{B_{s(d)}}
f_{B_{s(d)}}^2\hat B_{B_{s(d)}}\, ,
\ea
where $x_t=m_t^2/M_W^2$, $\eta_2^B$ is a perturbative QCD 
correction factor, $S_0(x_t)$ is the Inami-Lim 
function and the products $f_{B_{s(d)}}^2\hat B_{B_{s(d)}}$ 
parametrize the hadronic matrix elements in the effective theory with 
$f_{B_{s(d)}}$ the $B_{s(d)}^0$ decay constants and 
$\hat B_{B_{s(d)}}$ the (renormalization group invariant)  
bag parameters. The hadronic matrix elements can be calculated in
lattice QCD. Our current knowledge of them limits the accuracy with 
which the CKM matrix elements appearing in Eqn.~(\ref{SMMsd}) 
can be determined from the experimental measurements of 
$\Delta M_{s(d)}$. The goal of our project is to calculate all the 
hadronic matrix elements which are relevant for the mass and width 
differences in the $B_{s(d)}^0$ systems in unquenched lattice QCD 
at the few percent level. 



Many of the uncertainties that affect the theoretical calculation
of the decay constants and bag parameters cancel totally or partially 
if one takes the ratio $\xi^2=f_{B_s}^2 B_{B_s}/f_{B_d}^2 B_{B_d}$. 
Hence, this ratio and therefore the combination of CKM matrix elements 
related to it by Eqn.~(\ref{SMMsd}) can be determined with a significantly 
smaller error than the individual matrix elements. This is a crucial
ingredient in the unitarity triangle analysis.
In these proceedings we report our preliminary results for 
the determination of $\xi$, as well as for the quantities 
$f_{B_q}^2 B_{B_q}$.

Other work on this subject using $2+1$ lattice QCD
methods can be found in \cite{otherB0}.

\section{Operators, actions and matching calculation} \label{operandact}

The whole set of operators whose matrix elements are needed to 
determine the $B^0_{s(d)}$ mixing parameters are
\ba\label{operators}
&Q^{s(d)}=\left[\bar b^i \gamma_\mu(1-\gamma_5)s^i(d^i)\right]\,
\left[\bar b^j\gamma^\mu(1-\gamma_5)s^j(d^j)\right] \, ,&
\nonumber\\  
&Q^{s(d)}_S=\left[\bar b^i(1-\gamma_5)s^i(d^i)\right]\,
\left[\bar b^j(1-\gamma_5)s^j(d^j)\right] \, ,&
\nonumber\\
&Q^{s(d)}_3=\left[\bar b^i(1-\gamma_5)s^j(d^j)\right]\,
\left[\bar b^j(1-\gamma_5)s^i(d^i)\right]\, ,&
\ea
where $i,j$ are color indices. 
In these proceedings we focus on the results for the first two pairs 
of operators, enough to determine $\Delta M_{s(d)}$, and leave 
the study of the third pair, needed for an improved determination of  
$\Delta \Gamma_{s(d)}$, for a forthcoming publication \cite{EGE07}.  

We use the Fermilab action \cite{Fermilabaction} for the $b$ valence 
quarks and 
the Asqtad action \cite{Asqtadaction}, for the light sea and valence 
quarks, $u$, $d$ and $s$. The Fermilab action 
has errors starting at 
$\order(\alpha_s\Lambda_{QCD}/M)$ and $O((\Lambda_{QCD}/M)^2)$, 
while the errors of the Asqtad action are $\order(\alpha_s a^2, a^4)$. 

The products $f_{B_{s(d)}}^2 B_{B_{s(d)}}^{\overline{MS}}$ in 
Eqn.~(\ref{SMMsd}) parametrize the matrix elements by 
\ba\label{relation}
\langle \bar B^0_s\vert Q^{s(d)} \vert B^0_s\rangle^{\overline{MS}}(\mu) = 
\frac{8}{3}M^2_{B_{s(d)}}f^2_{B_{s(d)}}B_{B_{s(d)}}^{\overline{MS}}(\mu)\, .
\ea
The lattice matrix elements  
$\langle \bar B^0_{s(d)}\vert Q^{s(d)} \vert B^0_{s(d)}\rangle^{\rm lat}$ 
determine $f_{B_{s(d)}}^2 B_{B_{s(d)}}$ at tree 
level. Beyond tree-level, the operators $Q^{s(d)}$, mix with $Q_S^{s(d)}$ 
both on the lattice and in the continuum.  
Including one-loop corrections, the renormalized matrix element is
given by 
\ba\label{matching}
\frac{a^3}{2 M_{B_{s(d)}} } \; \langle Q^{s(d)} 
\rangle ^{\overline{MS}} (\mu)= 
[ 1 + \alpha_s \cdot \rho_{LL}(\mu,m_b) ] 
\langle Q^{s(d)} \rangle^{\rm lat} (a) + \alpha_s \cdot \rho_{LS}(\mu,m_b)
\langle Q_S^{s(d)} \rangle^{\rm lat} (a)\, .
\ea
The $\order\left(\frac{\Lambda_{QCD}}{M}\right)$ improvement is 
implemented by a rotation of the $b$ quark as explained in 
\cite{Fermilabaction}, so the perturbative matching errors start at
$\order(\alpha_s^2,\alpha_s \Lambda/aM)$. The matching 
coefficients $\rho_{LL}$ and $\rho_{LS}$ are the differences between 
the continuum $\overline{MS}$ and lattice renormalization coefficients 
calculated at one-loop order. We have calculated 
these coefficients for the same choice of lattice actions as used in the 
numerical simulations. We have checked that our results have
the correct infrared behavior, that they are correct
in the massless limit, and that they are gauge invariant. 
However, our results for the matching coefficients are still preliminary,
because not all diagrams have been independently checked. 

The optimal value of the strong coupling constant to be used in 
Eqn.~(\ref{matching}) is the V-scheme coupling 
$\alpha_V(q^*)$ \cite{LM93}  where $q^*$ is defined in \cite{LM93,BLM83}. 
In the absence of a calculation of $q^*$ for 
the specific processes we are studying, we choose $q^*=2/a$, very close 
to the $q^*$ calculated for heavy-light currents. The specific values 
for $\alpha_s$ we use are given in Table~\ref{simulation}.

\section{Simulation details}

The matrix elements needed to determine both $f_{B_q}^2B_{B_q}$ 
and $B_{B_q}$, are extracted from the following three-point and two-point 
functions
\ba\label{correlators}
&C_O(t_1,t_2)=
\displaystyle\sum_{\vec{x},\vec{y}}\langle \bar b(\vec{x},t_1) 
\gamma_5 q(\vec{x},t_1)\vert O(0) \vert \bar b(\vec{y},t_2) 
\gamma_5 q(\vec{y},t_2)\rangle ,
&\nonumber\\
&C_Z(t)=\displaystyle\sum_{\vec{x}}\langle \bar b(\vec{x},t) 
\gamma_5 q(\vec{x},t)\bar q(0)\gamma_5 b(0)\rangle, \quad
C_{A_4}(t)=\displaystyle\sum_{\vec{x}}\langle \bar b(\vec{x},t) 
\gamma_0 \gamma_5 q(\vec{x},t)\bar q(0)\gamma_5 b(0)\rangle\, ,& 
\ea
where the operator $O$ is any $Q^{s(d)}$ or $Q_S^{s(d)}$ defined in 
Eqn.~(\ref{operators}). The $B$ meson operators 
are smeared at the sink with a 1S onium wavefunction. All the correlation 
functions in Eqn.~(\ref{correlators}) are calculated using the open meson 
propagator method described in \cite{EEP06}.

We have performed these calculations on the MILC coarse lattices 
($a=0.12\,fm$) with 2+1 sea quarks and for three different sea light quark 
masses. The strange sea quark mass is always set to $0.050$. 
The light sea masses, $m_l\equiv m_{light}^{sea}$, 
number of configurations and other simulation details are collected 
in Table~\ref{simulation}. 
The mass of the bottom quark is fixed to its physical value, while 
for each sea quark mass we determine the different matrix elements 
for six different values of the light valence quark mass in a generic 
meson $B^0_q$, $m_q=0.0415,0.03,0.020,0.010,0.007,0.005$. 
We use $m_q=0.0415$ for  the valence strange mass in our simulations.
It is close to the physical strange quark mass, $m_s^{phys.}=0.036$ 
\cite{masa_s}. The matrix elements of the operators are extracted from 
simultaneous fits of three-point and two-point functions using 
Bayesian statistics. 
\begin{center}
\begin{table}[t]
\begin{center}
\begin{tabular}{c c c c c c}\hline\hline
$m_l/m_s^{sea}$ & Volume & $N_{confs}$ & $a^{-1}(\GeV)$ 
& $\alpha_s=\alpha_V(2/a)$ & $N_{sources}$\\
\hline
0.020/0.050 & $20^3\times 64$ & 460 & 1.605(29) & 0.31 & 4\\
0.010/0.050 & $20^3\times 64$ & 590 & 1.596(30) & 0.31 & 4\\
0.007/0.050 & $20^3\times 64$ & 890 & 1.622(32) & 0.31 & 4\\
\hline
\hline
\end{tabular}
\end{center}
\caption{Simulation parameters and $\alpha_s$ used in the matching 
with the continuum. $m_l$ is the light sea quark mass.\label{simulation}}
\end{table}
\end{center}

\section{Results}

\begin{figure}[th]
\begin{center}
\vspace*{-0.5cm}
\includegraphics[width=0.75\textwidth,angle=-90]{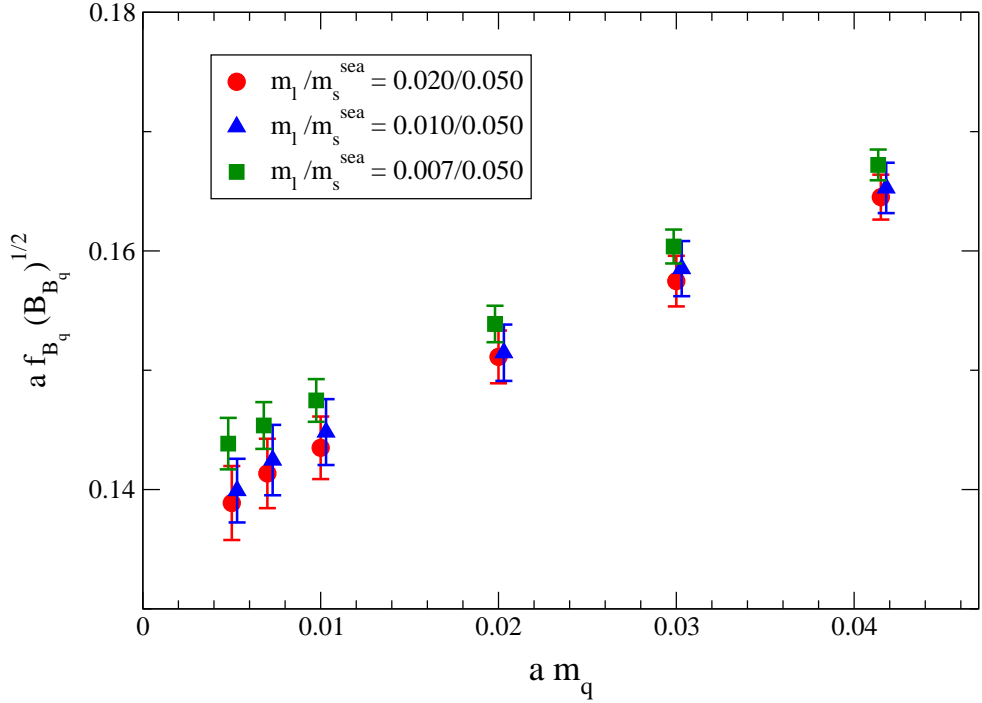}
\end{center}
\caption{$f_{B_q}\sqrt{B_{B_q}^{\overline{MS}}(m_b)}$ in lattice units. 
The different symbols and colors correspond 
to different values of the sea light quark masses, $m_l$.\label{fBBB}}
\end{figure}

The results in Fig.~\ref{fBBB} show 
$f_{B_q}\sqrt{B_{B_q}^{\overline{MS}}(m_b)}$ in lattice units as a 
function of the light valence mass $a\,m_q$. The errors shown are statistical 
errors only; the analysis of the systematic errors is not yet complete. 

The statistical errors range between $1-3\%$. Some conclusions can be 
already extracted from this plot. The light sea quark mass dependence of
$f_{B_q} \sqrt{B_{B_q}}$ is small compared to the statistical errors. 
The dependence on the light valence quark mass, however, is noticeable 
within statistics. In order to get a value for 
$f_{B_s}\sqrt{B_{B_s}^{\overline{MS}}(m_b)}$, since the $s$ valence 
quark mass we are using is slightly larger than the physical one, 
we need an interpolation in the $s$ valence quark mass together 
with a chiral extrapolation to the physical sea quark masses. 
To determine $f_{B_d}\sqrt{B_{B_d}^{\overline{MS}}(m_b)}$ we need 
extrapolations in both the $d$ valence quark mass and the sea quark masses. 
This is in progress. 

\begin{figure}[th]
\begin{center}
\vspace*{-0.3cm}
  \includegraphics[width=0.75\textwidth,angle=270]{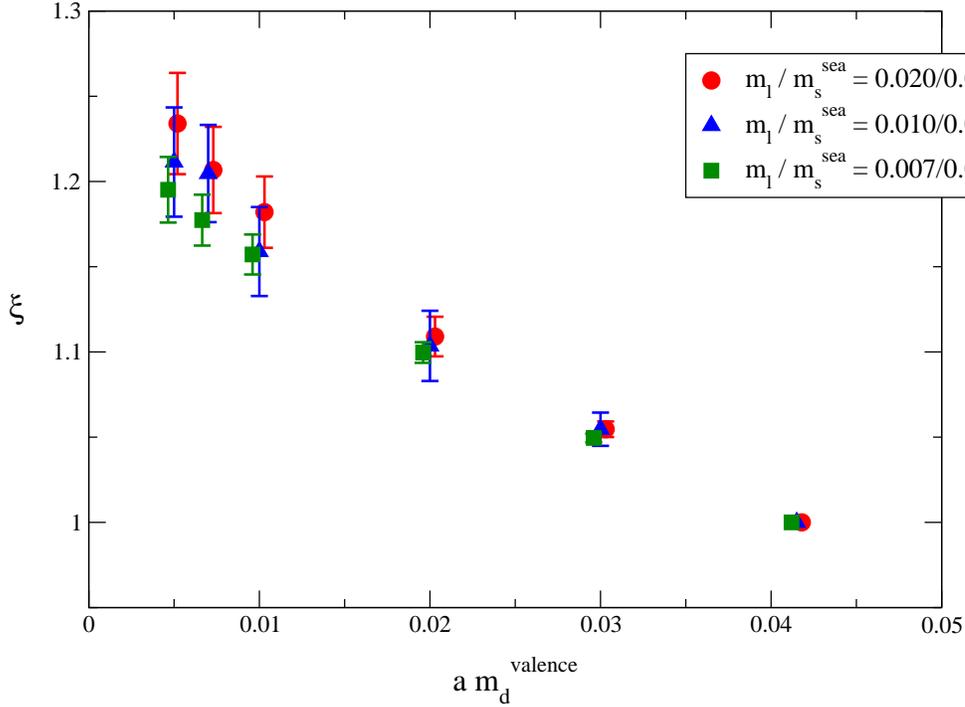}
\end{center}
\caption{$\xi$ as a function of the valence $d$ mass for 
three different values of the light sea quark masses.\label{xi}}
\end{figure}

In Fig.~\ref{xi} we plot the ratio $\xi=f_{B_s}\sqrt{B_{B_s}}
/f_{B_d}\sqrt{B_{B_d}}$ as a function of the $d$ valence quark mass 
in the denominator. Again, our results are preliminary for the same 
reasons as mentioned before and the errors are only statistical. 
Most of the systematic errors cancel in the ratio, but not those 
associated with the chiral extrapolation in the light valence quark mass. 

\subsection{Chiral extrapolation}

\label{chpt}

The continuum chiral expansion of the hadronic matrix element 
$\left<\bar{B}_q|Q|B_q\right>$ at NLO in (partially quenched) heavy 
meson chiral perturbation theory (HMChPT) is given by \cite{Detmold}
\begin{equation}
\label{xpt}
\left<\bar{B}_q|Q|B_q\right>=\beta(1+w(T_q+W_q+S_q))
+c_0m_q+c_1(m_U+m_D+m_S)
\end{equation}
where $m_U,m_D,m_S$ are the sea quark masses and $m_q$ the light 
valence quark mass. $\beta$, $w$, $c_0$ and $c_1$ are low energy 
constants (LECs) to be determined from the fits. The functions $T_q$, 
$W_q$, and $S_q$ contain the chiral logs and correspond to 
tadpole-, wave-function, and sunset-type contributions, respectively. 

The effects of $\order (a^2)$ taste changing interactions can be included
in Eqn.~(\ref{xpt}) using staggered chiral perturbation theory 
(SChPT). 
In that case, the chiral log functions are modified to depend on the 
masses of the different taste multiplets. Explicit expressions from SChPT 
for heavy-light bilinear quantities can be found in \cite{Aubin,Aubin2}. 
Similar terms are expected to contribute to our four-quark operators.
The modified chiral logs contain other fixed LEC's, most of which are 
already determined to a high degree of certainty \cite{Aubin2}. 
The logs also contain constants from heavy quark effective theory, 
in particular the mass splitting between the vector and pseudoscalar 
heavy mesons, $\Delta*=M_B*-M_B$, and the mass splittings between the 
pseudoscalar heavy mesons containing different valence and sea light quarks,  
$\delta_{qr}=M_{B_r}-M_{B_q}$.  These HQET constants can be determined 
directly from the two-point function fits and used as input into the chiral 
fits, with the experimental values then used in the extrapolation.  

We are still in the process of determining the exact SChPT form of
Eqn.~(\ref{xpt}). Once the functional SChPT form of Eqn.~(\ref{xpt}) is 
completely determined,
we plan to use it to simultaneously fit it to our lattice data points 
for all sea and valence quark masses, and to determine the unknown LEC's 
in the process. For the systematic error analysis, we plan to study the 
effects of changing the SChPT form, for example by adding NNLO analytic terms, 
and the effects of allowing the more poorly known fixed parameters to 
vary. Our physical results will then be obtained by turning
the taste-violations off, and extrapolating (interpolating) to the
physical light (strange) sea and valence quark masses. 

We will quote results for $\xi$, $f_{B_s}\sqrt{B_{B_s}}$, 
$f_{B_d}\sqrt{B_{B_d}}$, $B_{B_s}$, and $B_{B_d}$ when this step is 
completed.  We expect light quark discretization effects to be an important 
source of uncertainty until we calculate the three-point correlators at 
several lattice spacings and use these in the chiral fits.  


\section{Summary and future work}

We have presented preliminary results for $f_{B_q}\sqrt{B_{B_q}}$ 
for six different values of $m_q$ as well as for the ratio $\xi$ 
with five different values of $m_d^{valence}$. Our analysis on three 
ensembles with different sea light quark masses gives statistical 
errors between $1-3\%$ for $f_{B_q}\sqrt{B_{B_q}}$ and $1-2\%$ for 
the ratio $\xi$. The systematic error analysis is 
in progress, see Table~\ref{errors}.
\begin{table}
\begin{center}
\begin{tabular}{ccc}
\hline
\hline
&$f_{B_q}\sqrt{B_{B_q}}$ & $\xi$\\
\hline
statistics & $1-3$ & $1-2$ \\
scale($a^{-1}$) & $0.9$ & $0$ \\
Higher order matching & $\sim 4.5$ & 
\hspace*{-0.6cm}cancel to a large extent\\
Heavy quark discret. &   $2-3$ & $<0.5$ \\
Light quark discret. + $\chi$PT fits &\multicolumn{2}{c}
{ Work in progress}\\
\hline
\hline
\end{tabular}
\caption{Error budget for $f_{B_q}\sqrt{B_{B_q}}$ and $\xi$
in percent.
\label{errors}}
\end{center}
\end{table}

Two important sources of error in the calculation of 
$f_{B_q}\sqrt{B_{B_q}}$, the matching uncertainties and heavy quark 
discretization errors, are expected to cancel to a large extent 
when taking the ratio. 
We have already checked that the difference between the tree level 
and the one-loop results for $\xi$ with our preliminary results for 
the renormalization constants is less than $0.5\%$. The higher order 
matching errors in Table~\ref{errors} have been 
naively estimated as being $\order{\left(1\times\alpha_s^2\right)}$ 
for the coarse lattice. Heavy quark discretization effects in the table 
are estimated by power counting \cite{fdprl}. 

We are in the process of generating lattice data on the coarse lattice 
with a smaller light sea quark mass, $m_l=0.005$, which will further
constrain the chiral extrapolations. Better fitting approaches and 
different smearings that could reduce statistical errors further are 
also being investigated. Results from these improvements will be 
presented in a future publication \cite{EGE07}. We also plan to 
present results for the decay width differences 
$\Delta \Gamma_s$ and $\Delta \Gamma_d$, for which we have already 
calculated all the hadronic matrix elements needed. 
Finally, we plan to improve this analysis by repeating this 
calculation at other lattice spacings to study the discretization
errors in detail. Simulations on finer lattices will reduce both
discretization and perturbative matching errors.

\acknowledgments

The numerical simulations for this work were carried out on the 
Fermilab lattice QCD clusters, which are a computing resource 
of the USQCD collaboration and are funded by the DOE. We are 
grateful to the Fermilab Computing Divisions for operating and 
maintaining the clusters. This work was supported in part by the DOE 
under grant no. DE-FG02-91ER40677 and by the Junta de Andaluc\'{\i}a 
[P05-FQM-437 and P06-TIC-02302].

\end{document}